# New aspects in the Bragg Glass-Disordered phase transition: an analysis based on the 3$^{rd}$ harmonics of the AC magnetic susceptibility

M. G. Adesso[1], R. Flukiger[2], T. Giamarchi[2], W. Goldacker[3], H. Küpfer[3], S. Pace[1], M. Polichetti[1], D. Uglietti[2]

1. University of Salerno & Laboratorio Regionale Supermat CNR-INFM, Via S. Allende 84081 Baronissi (Salerno), Italy
2. DPMC-MaNEP, University of Geneva, 24 Quai Ernest Ansermet, 1211 Geneva 4, Switzerland
3. Forschungszentrum Karlsruhe, Institut fur Festkorperphysik, Postfach 3640, 76021 Karlsruhe, Germany

We analyse the phase transition between the Bragg Glass and the Disordered phase in the vortex lattice in type-II superconductors, both by analytical computations and experimental investigations. It is known that if the Peak Effect can be detected, a Bragg Glass/Disordered phase transition takes place. We show that, in some conditions, this transition can occur without the observation of the Peak Effect Phenomenon. We introduce a method based on the 3$^{rd}$ harmonics of the AC magnetic susceptibility to detect the transition also in these cases. Using this method, we obtain an experimental confirmation of the theoretical predictions on sphere shaped $V_3Si$ single crystals, in the high fields/low temperatures range too, where previous experimental studies failed to detect the Bragg/Disordered phase transition.

Understanding the statics and dynamics of the Abrikosov vortex lattice and the phase transitions in vortex matter plays a fundamental role, not only for the superconducting materials, but also for crystals and glassy systems, the vortex lattice being a model system for a general analysis of phase transitions in presence of quenched disorder [1-4].
A universal field-temperature (H-T) phase diagram has been proposed for all type-II superconductors, being characterized by a phase transition in the vortex matter between a disordered phase and the Bragg Glass (BrG) phase, where the lattice still survives in spite of the presence of disorder, generated by the pinning centres [5, 6]. The Bragg Glass phase has been directly observed by neutron diffraction experiments [7]. The detection of the Peak Effect [8], i.e. a local maximum in the critical current density ($J_c$) as a function of both the temperature and the magnetic field [9], is a direct evidence that a BrG/Disordered phase transition is occurring [5]. Different experimental techniques have been used to detect the Peak Effect [4, 6-13] and, consequently, to get information about the existence of the Bragg Glass/Disordered phase transition. In particular, one of these techniques is based on the measurements of the real part of the 1$^{st}$ harmonics of the AC magnetic susceptibility ($\chi_1^{'}$), either as a function of the temperature (T) or as a function of the DC field (H): the evidence of a dip in $\chi_1^{'}$ indicates that the capability of the superconductor to exclude the magnetic flux is not monotonic with temperature and/or magnetic field and it is directly associated to the Peak Effect [13]. A new method based on the measurements of the 3rd harmonics of the AC magnetic susceptibility has been recently introduced to detect the BrG/Disordered phase transition in $Nb_3Sn$ [14, 15].

In this letter we show that the BrG/Disordered phase transition is not strictly related to the presence of the Peak Effect but still exists in the low temperature/high field region, where the Peak Effect is not observed. An experimental confirmation of this result has been obtained on $V_3Si$ single crystals. Indeed by Peak Effect measurements, the BrG/Disordered phase transition was previously observed in the samples analysed here, at fields up to 7 T, [16, 17]. By measurements of 3rd harmonics, we succeeded here to measure the BrG/Disordered phase transition at higher fields, where other methods failed.

The Bragg Glass and the disordered phase are characterized by different voltage-current characteristics [5]. In particular, the distorted Abrikosov vortex lattice is collectively pinned in the Bragg Glass phase, while the disordered phase is described by a single vortex strong pinning model [5]. Thus a different temperature dependence of both critical current density ($J_c$) and pinning potential ($U_p$) characterizes the two vortex phases. In general, $J_c$ in the disordered phase is higher or equal to $J_c$ in the Bragg Glass phase, whereas the pinning potential is larger in the Bragg Glass with respect to the Disordered phase. In the present computations, for the Bragg Glass phase the Collective Flux Creep [18] was chosen as Collective pinning model, which corresponds to the following temperature dependence of the critical current density:

$$J_{cB}(t) = J_{cB0}(1-t^2)^2, \qquad (1)$$

with $t=T/T_c$. In the disordered phase, the Ginzburg-Landau model [19], which is a single pinning model, has been used:

$$J_{cD} = J_{cD0}\frac{(1-t^2)}{(1+t^2)}. \qquad (2)$$

At a fixed DC field, the BrG/Disordered phase transition, at a specific $t$ value, corresponds to a jump from one $J_c$ curve (solid line in Fig. 1) to the other (dotted line), with a possible consequent increase of the critical current (i.e. the Peak Effect phenomenon).

In Fig. 1.a, the critical current density as a function of the reduced temperature $t$ is shown, as computed both in the disordered and BrG phase, by using two different values of the critical current density at zero $T$, in particular: $J_{cD0}=2\times10^8$ A/m$^2$, $J_{cB0}=1\times10^8$ A/m$^2$. Under these conditions, the BrG/Disordered phase transition is always associated to the Peak effect, at all the temperatures.

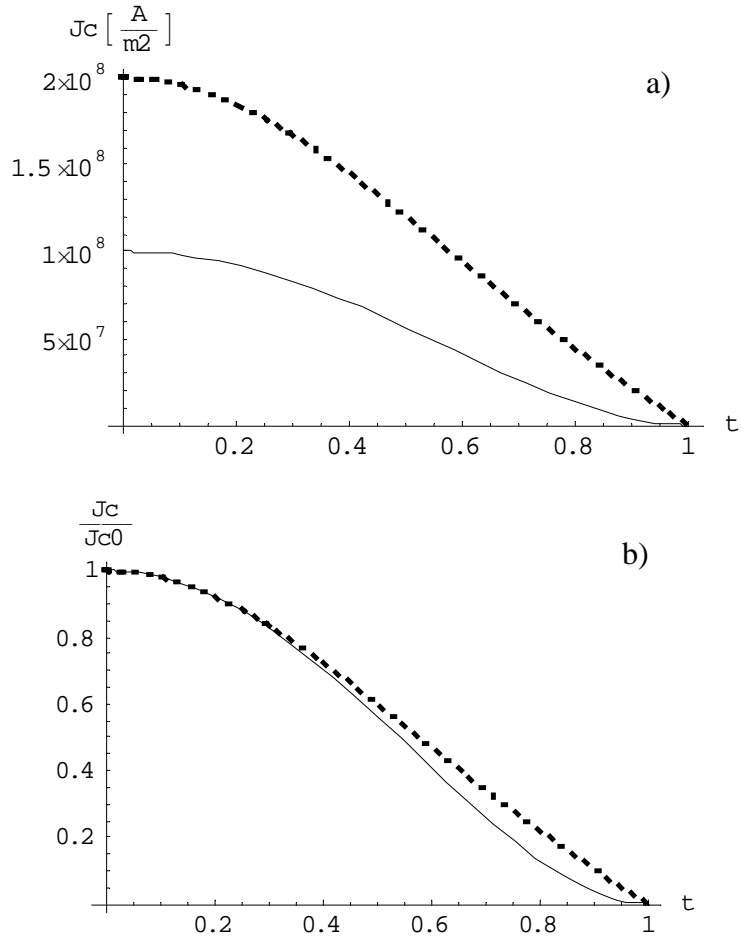

**Fig. 1** Behavior of the critical current density as a function of the reduced temperature in the Disordered phase (dashed line) and in the Bragg Glass phase (solid line), computed by using two different pinning models. The value of the critical current density at zero temperature was chosen either different (a) or equal (b) in the two phases.

In Fig. 1.b, the critical current curves obtained with the same critical current density value at zero temperature in both phases ($J_{cB0}=J_{cD0}=J_{c0}=10^8 A/m^2$) are plotted. Contrarily to the previous case, at low temperatures the BrG/Disordered phase transition is not associated to the Peak Effect anymore.

From the analysis of these curves, we deduce that the BrG/Disordered transition can also occur without being revealed by the Peak Effect phenomenon, if the critical currents in the two phases are close enough. However, we show that signatures of the transition still exist in the higher harmonics of the AC magnetic susceptibility.

By using these pinning models, we analytically computed the 1$^{st}$ and 3$^{rd}$ harmonics of the AC magnetic susceptibility in presence of a BrG/Disordered phase transition. The harmonics have been obtained by fixing a phase transition temperature ($T_p$) and imposing the relation (1) for temperatures lower than $T_p$ and the relation (2) at higher temperatures. In these computations, the critical state Bean model has been used [20, 21], thus neglecting the vortex dynamics [22]. By numerical solutions of diffusion equation of the magnetic induction [23], a more detailed analysis of the vortex dynamical effects on the harmonics and on the H-T phase diagram, with further confirmations of the interpretation here presented, will be discussed elsewhere [24].

In Fig. 2, the real part of the 1st and 3rd harmonics of the AC magnetic susceptibility is reported as a function of the reduced temperature, by supposing that the BrG/Disordered phase transition is occurring at high temperature ($T_p/T_c = 0.85$). The curves in Fig. 2.a are obtained with different $J_{c0}$ values, whereas those in Fig. 2.b are calculated with the same $J_{c0}$ for both phases. In Fig. 3, the corresponding harmonics are plotted, as computed by supposing a low transition temperature ($T_p/T_c = 0.55$).

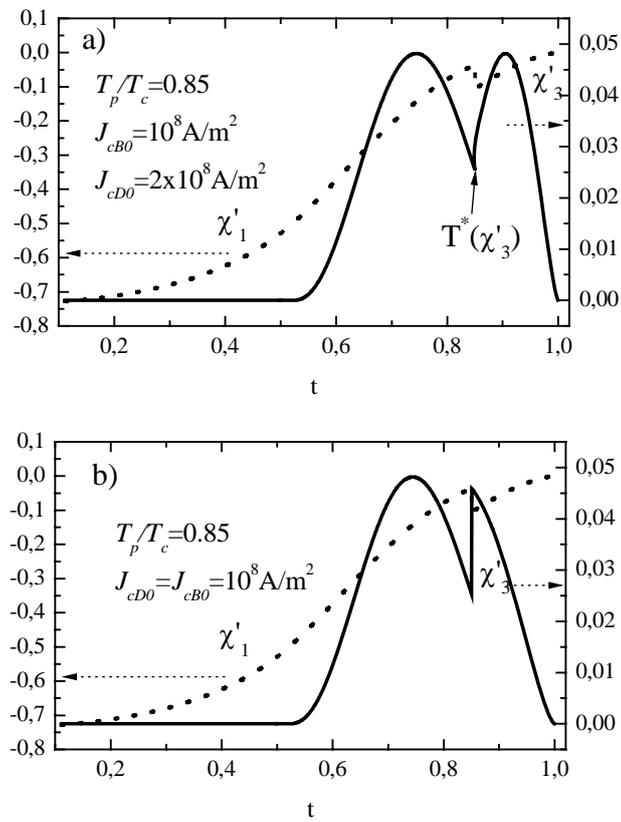

**Fig. 2** Real part of the 1st and 3rd harmonics as a function of the reduced temperature, computed in the Bean model framework, by supposing a Bragg/Disordered phase transition at a temperature $T_p/T_c=0.85$, in the following cases: a) $J_{cD0}=2\times10^8$ A/m$^2$, $J_{cB0}=1\times10^8$ A/m$^2$; b) $J_{cB0}=J_{cD0}=J_{c0}=10^8$ A/m$^2$.

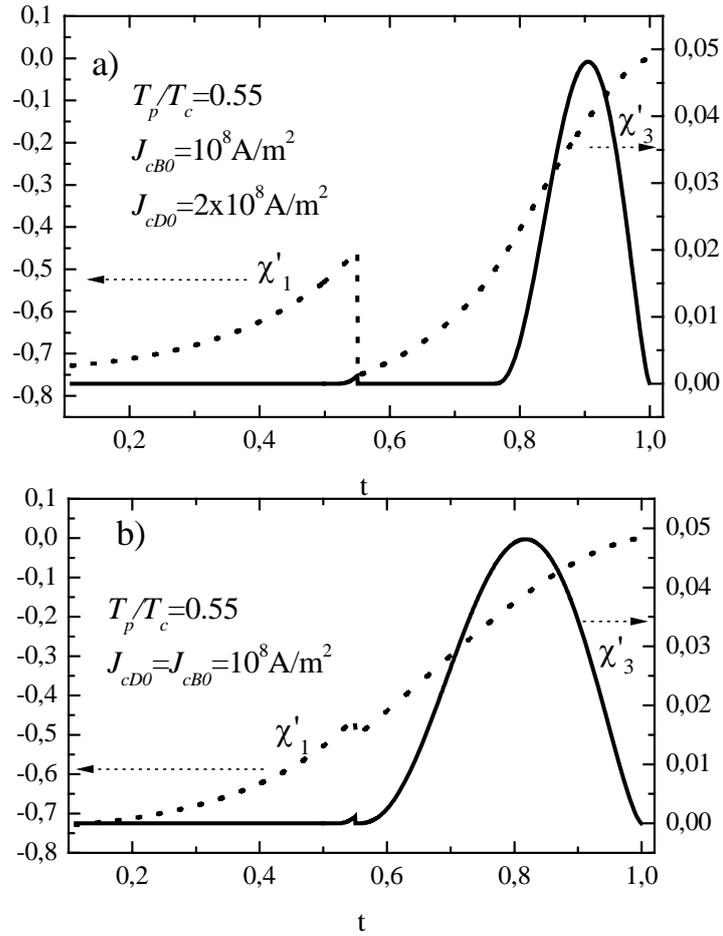

**Fig. 3** Real part of the 1st and 3rd harmonics as a function of the reduced temperature, computed by fixing a Bragg/Disordered phase transition at low temperatures ($T_p/T_c$=0.55), in the following cases: a) $J_{cD0}$=2x10$^8$ A/m$^2$, $J_{cB0}$=1x10$^8$ A/m$^2$; b) $J_{cB0}$=$J_{cD0}$=$J_{c0}$=10$^8$A/m$^2$.

If the BrG/Disordered phase transition is occurring at high temperatures, the dip in $\chi_1'$, which identifies the Peak Effect, can be always observed (Fig. 2), regardless of the $J_{c0}$ values in both phases. Correspondingly, the real part of the 3rd harmonics, $\chi_3'$, is characterized by two positive peaks, one near $T_c$ and the other at lower temperatures. The temperature $T^*(\chi_3')$ (i.e. the lowest temperature of the positive peak near $T_c$ in Fig. 2), and the dip temperature in the real part of the 1st harmonics both correspond to the BrG/Disordered phase transition temperature, $T_p$.

On the contrary, if the BrG/Disordered phase transition is occurring at low temperatures (see Fig. 3), two different behaviours can be observed, depending on the $J_{c0}$ values. In particular, if the two phases are characterized by the same $J_{c0}$ value, the Peak Effect is just slightly visible. Correspondingly, (see Fig. 3.b) the positive peak at low temperatures in the real part of the 3rd harmonics decreases more and more. Nevertheless, the BrG/Disordered phase transition can be detected by $T^*(\chi_3')$, which always corresponds to the transition temperature $T_p$. Nevertheless, if the values of $J_{c0}$ are different in the two phases (Fig. 3.a), a Peak Effect can be always detected at a temperature $T_p$, corresponding to a low but still evident positive peak in the real part of the 3rd harmonics of the AC magnetic susceptibility.

This analysis shows that if the Bragg Glass phase and the disordered one are characterized by the same $J_{c0}$ value the Peak Effect is valuable to detect the phase transition between the two phases only at high temperatures (and correspondingly low DC fields) range. At low temperatures it fails, whereas the use of the 3rd harmonics allows in any case to detect this phase transition.

Moreover, by taking into account the vortex dynamical effects [22-24], the disordered phase is qualitatively characterized by a significant negative peak in the real part of the third harmonics, at temperatures lower than $T^*(\chi_3')$ whereas, in a first approximation, the Bragg Glass phase can be correctly described by the critical state model, corresponding to $\chi_3' = 0$ for $T < T^*(\chi_3')$. In this sense, the absence of negative values of $\chi_3'$ for temperatures lower than $T^*(\chi_3')$ is a further confirmation of the BrG/Disordered phase transition at the $T^*(\chi_3')$ temperature.

In order to verify the predictions mentioned above, we experimentally studied the magnetic response of spherical $V_3Si$ single crystals (diameter d=1,8 mm). More details about the analyzed sample can be found in Ref. [16]. In literature, a Peak Effect in the $V_3Si$ has been already observed [25]. In particular, Küpfer et al. [16] detected a BrG/Disordered phase transition in the same single crystal as analysed here, at reduced temperatures up to t=0.7 (correspondingly to a field H=7 T). Since it was not possible to detect the Peak Effect at lower temperatures/higher fields, the authors supposed that the Bragg Glass Phase extended without a transition up to the vortex liquid state [16], in disagreement with the theoretical predictions [4-6].

We performed measurements of both the 1st and 3rd harmonics of the AC magnetic susceptibility as a function of the temperature, at a fixed amplitude ($h_{AC}$ = 128 Oe) and frequency ($\nu$=107 Hz) of the AC magnetic field, at various DC fields up to 13 T, parallel to the AC field.

In Fig. 4, the real part of the 1st and 3rd harmonics of the AC magnetic susceptibility is reported as a function of the temperature, at various DC fields. A dip appears in the real part of the 1st harmonics (Fig. 4a), indicating that a Peak Effect is occurring [13] in the field range 1 T - 7 T, in agreement with previous magnetic measurements [16]. Nevertheless, the Peak Effect cannot be detected anymore at higher fields/lower temperatures. The corresponding 3rd harmonics are characterized by two main positive peaks. A peak at high temperatures has been detected at all field/temperature values, its height and width being almost constant with the DC field. Moreover, the height of the positive peak at lower temperatures decreases with $H$ and tends to disappear at high fields/low temperatures. The behaviour of the positive peaks in $\chi_3^{'}$ measured on V$_3$Si is analogous to the Nb$_3$Sn magnetic response [14, 15]. Moreover, a "small" negative peak can be also detected near $T_c$, with an almost constant height, due to dynamical effects [24].

The Field-Temperature phase diagram has been obtained by plotting the superconducting critical temperature $T_c$, the temperature $T_p$ corresponding to the dip in $\chi_1^{'}$, and the lowest temperature of the positive peak near $T_c$ in $\chi_3^{'}$, $T^*\left(\chi_3^{'}\right)$, at each applied DC field. In Fig. 5, the reduced field-temperature phase diagram ($h=H/H_{c2}$ vs $t=T/T_c$, where $H_{c2}$=20 T and $T_c$=17 K) is reported.

It is remarkable that $T^*\left(\chi_3^{'}\right)$ corresponds to $T_p$, at a reduced temperature up to 0.75, thus confirming the predictions reported above. It has to be noted that $T^*\left(\chi_3^{'}\right)$ can be measured at high fields/low temperatures too, where the Peak Effect is not observed.

Our analytical computations showed that, if $J_{c0}$ is the same in both phases (Bragg and Disordered), the detection of the BrG/Disordered phase transition at low temperatures is not possible via the Peak Effect: the experimental results reported here on V$_3$Si, represent an evidence of this case. Nevertheless, by using the innovative analysis of the 3$^{rd}$ harmonics, the phase transition between Bragg and Disordered phase has been detected in an extended range up to higher fields/lower temperatures, thus confirming the agreement with the theoretical predictions [5, 6].

As previously stated, a further confirmation of these results can be also given by the shape of the 3$^{rd}$ harmonics. Due to vortex dynamical effects, it is known that the disordered phase, in addition to the positive peak near $T_c$, is characterized by a negative peak at lower temperatures [13]. The "absence" of this negative peak in our experimental data, especially at high fields, confirms the presence of the Bragg Glass phase. A more detailed analysis of these data, based on numerical simulations including the BrG/Disordered phase transition and the vortex dynamics, will be discussed in [24].

Summarizing, in this letter we prove that the BrG/Disordered phase transition can occur without the presence of the Peak Effect phenomenon at low temperatures/high fields, if the critical current density at zero temperature is the same in both vortex phases. In this case, we show that the analysis of the 3$^{rd}$ harmonics is valuable to detect this phase transition. Moreover, an experimental confirmation of these results has been obtained on V$_3$Si single crystal samples. The presented tools lead to a much more complete picture of the vortex matter phase diagram and open the way for a better interpretation of some open questions reported in literature, like, for example, the presence of an eventual ending point on the Bragg Glass phase transition at 20 K in Bi-2212 [26-28].

The general applicability of our analysis to all type-II superconductors give us the opportunity to better investigate (both theoretically and experimentally) the universal field-temperature phase diagram for the entire class of these materials and to improve our knowledge in the phase transition framework.

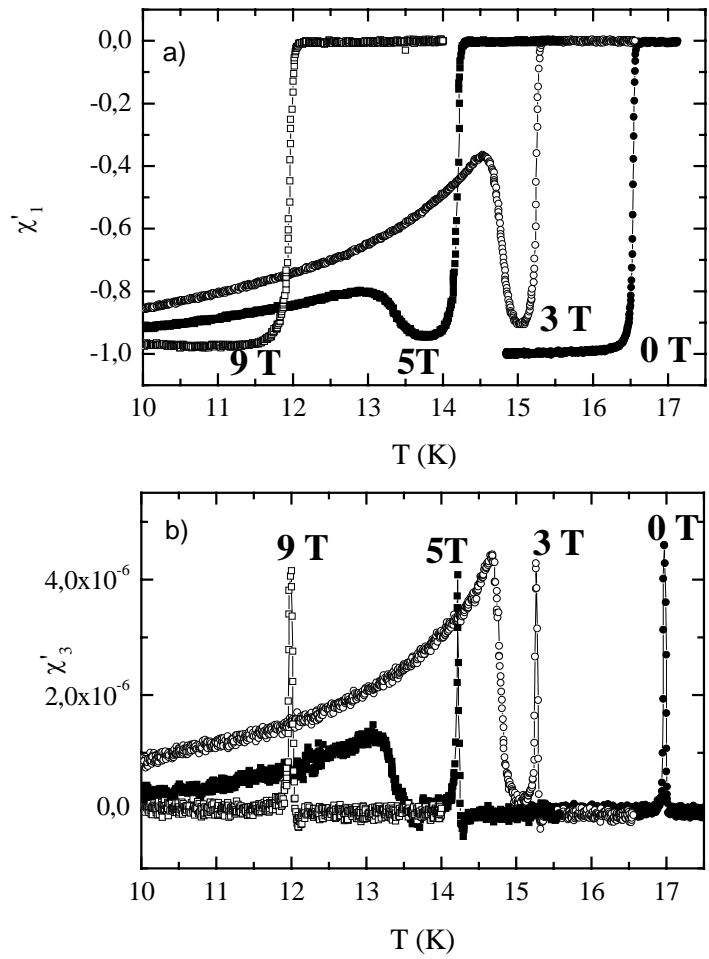

**Fig. 4** Real part of the 1$^{st}$ (a) and 3$^{rd}$ (b) harmonics of the AC magnetic susceptibility as a function of the temperature, measured on V$_3$Si single crystal, at various DC field.

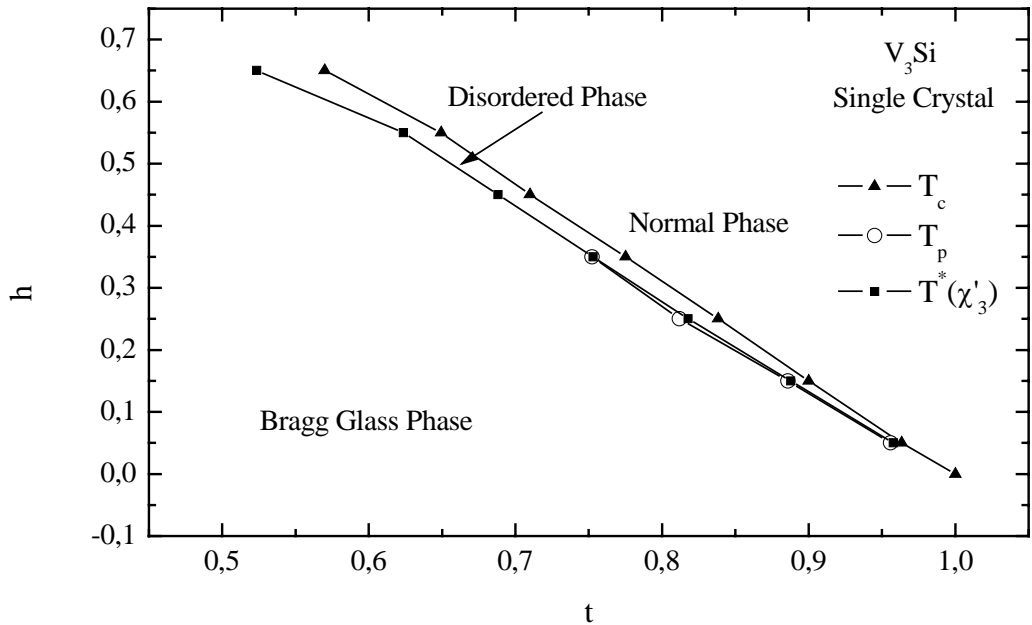

**Fig. 5** Reduced Field-Temperature phase diagram, measured on V$_3$Si sphere shaped single crystal, obtained by measurements of both the 1$^{st}$ and 3$^{rd}$ harmonics of the AC magnetic susceptibility.

*Acknowledgements*

The authors are really grateful to B. Seeber, R. Moudoux, and A. Ferreira for helping in the laboratory.

This work was supported in part by the FNS under MaNEP and division II.